\documentclass[aps,prl,superscriptaddress,showpacs,reprint,floatfix]{revtex4-2}

\usepackage{blindtext}
\usepackage{centernot}
\usepackage{graphicx}
\usepackage{amsmath,bbold}
\usepackage{times}
\usepackage{amssymb}
\usepackage{mathrsfs}
\usepackage{chemarr}
\usepackage[svgnames]{xcolor}
\usepackage{url}
\usepackage{version}
\definecolor{darkgreen}{RGB}{0,100,0}
\definecolor{magenta}{RGB}{255,0,255}

\usepackage{tikz}
\usetikzlibrary{decorations.pathmorphing}
\usetikzlibrary{arrows.meta}
\usetikzlibrary{calc}
\usepackage[percent]{overpic}
\usepackage{bm}
\definecolor{linkcolor}{rgb}{0,0,0.6}

\newcommand{\bp}{{\bf p}}
\newcommand{\bx}{{\bf x}}
\newcommand{\bq}{{\bf q}}
\newcommand{\br}{{\bf r}}

\newcommand{\cL}{\mathcal{L}}

\usepackage{lipsum}
\usetikzlibrary{patterns}

\usepackage[pdftex]{hyperref}
\graphicspath{{../}}
\begin{document}

\title{A cold tracer in a hot bath: in and out of time}
\title{A Cold Tracer in a Hot Bath: In and Out of Equilibrium}

\author{Amer Al-Hiyasat}
\affiliation{Department of Physics, Massachusetts Institute of Technology, Cambridge, Massachusetts 02139, USA}
\author{Sunghan Ro}
\affiliation{Department of Physics, Harvard University, Cambridge, Massachusetts 02138, USA}

\author{Julien Tailleur}
\affiliation{Department of Physics, Massachusetts Institute of Technology, Cambridge, Massachusetts 02139, USA}

\date{\today}
\begin{abstract}
We study the dynamics of a zero-temperature overdamped tracer in a
bath of Brownian particles. As the bath density is increased,
numerical simulations show the tracer to transition from an active
dynamics, characterized by boundary accumulation and ratchet currents,
to an effective equilibrium regime.
To account for this analytically, we eliminate the bath degrees of
freedom under the assumption of linear coupling to the tracer and show
convergence, in the large density limit, to an equilibrium dynamics at
the bath temperature.
We then develop a perturbation theory to characterize the tracer's
departure from equilibrium at large but finite bath densities,
revealing an intermediate time-reversible yet non-Boltzmann regime,
followed by a fully irreversible one. Finally, we show that when the
bath particles are connected as a lattice, mimicking a gel or a soft active solid, the cold
tracer drives the entire bath out of equilibrium, leading to a
long-ranged suppression of bath fluctuations.
\end{abstract}
\maketitle

Characterizing the dynamics of a tracer particle immersed in a bath is a fundamental problem in statistical mechanics. For passive, equilibrium fluids, considerable effort has been devoted to deriving effective tracer dynamics starting from microscopic Hamiltonian descriptions~\cite{feynman1963, fkm1965, caldeira1983,  mori1965, vankampen1986, zwanzig2001}. 
These studies conclude that equilibrium is contagious: if the bath
particles are in equilibrium, the tracer follows a Langevin equation
that leads to the Boltzmann distribution. More recently, attention has
shifted to the converse question of whether nonequilibrium is
contagious.  For the much-studied case of a passive tracer in a bath
of active
particles~\cite{wu2000,Loi2008,Underhill2008,Leptos2009,Dunkel2010,Kurtuldu2011,Mino2011,Zaid2011,wilson2011differential,Foffano2012,Mino2013,Kasyap2014,maggi2014generalized, Morozov2014,Thiffeault2015,Suma2016,Argun2016,Burkholder2017,
  Pietzonka2018,cui2018generalized,
  Chaki2018,Kanazawa2020,Knezevic2020,maes2020,Abbaspour2021,Reichert2021,
  granek2022, solon2022, santra2023dynamical,
  pelargonio2023generalized,sarkar2024harmonic}, the answer is
generally affirmative: the tracer exhibits anomalous transport
properties, violates appropriately defined fluctuation-dissipation
theorems (FDTs), and generates ratchet currents that can induce
spontaneous motion. The tracer thus inherits the active nature of the
bath.

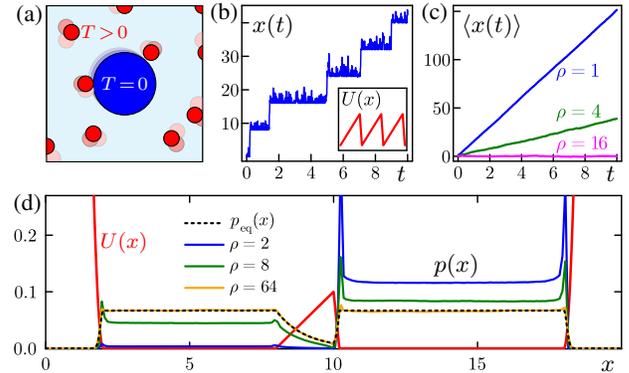
\begin{figure}
\raisebox{0.87em}{
\begin{tikzpicture}[scale=0.79]
    \newlength{\diagwidth}
    \setlength{\diagwidth}{0.304\linewidth}
    \filldraw[fill=cyan!10] (-0.5\diagwidth,-0.5\diagwidth) rectangle (0.5\diagwidth,0.5\diagwidth);
    \begin{scope}
    \clip (-0.5\diagwidth,-0.5\diagwidth) rectangle (0.5\diagwidth,0.5\diagwidth);
     \filldraw[fill=blue!20, draw=black!20] (-0.03\diagwidth,0.03\diagwidth) circle (0.2\diagwidth);
     \filldraw[fill=blue!40, draw=black!40] (-0.015\diagwidth,0.015\diagwidth) circle (0.2\diagwidth);
     \filldraw[fill=blue, draw=black] (0,0) circle (0.2\diagwidth);

    \foreach \x/\y in {0.17/0.2, -0.2/-0.32, 0.3/-0.35, -0.34/0.33, -0.25/0.0, -0.5/-0.4, 0./0.5, 0.45/-0.2, 0.52/0.2, 0.5/0.5} {
    \filldraw[fill=red!20, draw=black!20] (\x*\diagwidth-\y*0.1*\diagwidth,\y*\diagwidth+\x*0.15\diagwidth) circle (0.05\diagwidth);
     \filldraw[fill=red!40, draw=black!40] (\x*\diagwidth+\x*0.1*\diagwidth,\y*\diagwidth+\y*0.1*\diagwidth) circle (0.05\diagwidth);
     \filldraw[fill=red, draw=black] (\x*\diagwidth,\y*\diagwidth) circle (0.05\diagwidth);}
     \draw(-0.5\diagwidth,-0.5\diagwidth) rectangle (0.5\diagwidth,0.5\diagwidth);
    \node[text=red] at (-0.13\diagwidth,0.326\diagwidth) {\scriptsize$T\!>\!0$};
     \node[text=white, ] at (0,0) {\scriptsize$T\!=\!0$};
     \end{scope}
     \node at (-1.6,1.16) {(a)};
\end{tikzpicture}}\hspace{0.3em}
\begin{overpic}[scale=0.54, trim={0.4cm 0.5cm 0.3cm 0.4cm}, clip]{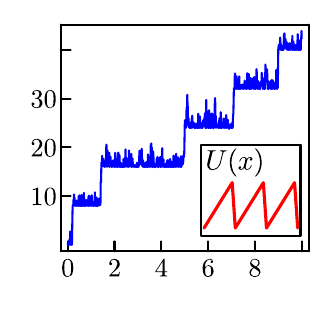} \put(-3,81.5){(b)} \put(90,1){$t$} 
\put(18,75){$x(t)$}
\end{overpic}\hspace{0.1em}
\begin{overpic}[scale=0.54, trim={0.4cm 0.5cm 0.3cm 0.4cm}, clip]{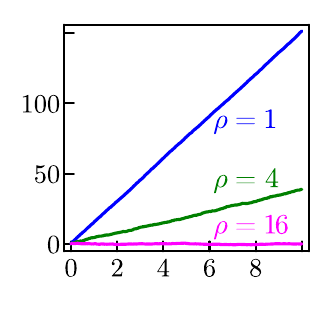} \put(-2,81.5){(c)} \put(90,1){$t$} 
\put(18,75){$\langle x(t) \rangle$}
\end{overpic} 

\hspace{.7em}\begin{overpic}[percent, scale=0.54, trim={0.6cm 0.2cm 0.3cm 0.1cm},clip]{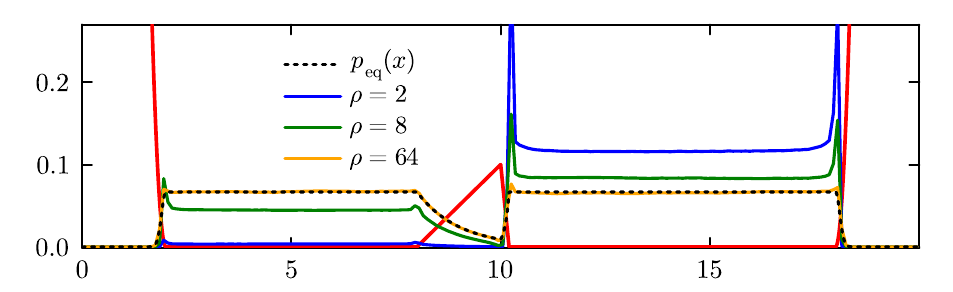} 
\put(0,26){(d)} 
\put(68,16){$p(x)$} 
\put(14,20){\footnotesize \color{red}{$U(x)$}}
\put(95, 0){$x$}
\end{overpic}

\caption{{\bf (a)} A zero-temperature tracer in a bath of $N$ Brownian particles at $T>0$, with short-ranged repulsive interactions between the tracer and bath particles.  {\bf (b)} Tracer position in a ratchet potential $U(x)$ of period $L$ in units of tracer radius. {\bf (c)} The mean position $\langle x(t)\rangle$ scales as $t \langle \dot x \rangle$, demonstrating the existence of a ratchet current $\langle \dot x \rangle$ that decreases with the bath density $\rho=N/L^d$. {\bf (d)} An external potential $U(x)$ models both confining walls at the system's ends and an asymmetric obstacle in the bulk. The stationary probability density $p(x)$ of the tracer shows that the rectification and boundary accumulation observed for finite bath densities disappear as $\rho\to\infty$.} 
\label{fig:fig1}
\end{figure}

An interesting case, intermediate between the equilibrium and active
baths described above, is that of a cold tracer in a bath of hot
Brownian particles
(Fig. \ref{fig:fig1}a)~\cite{ilkerjoanny2021,grosberg2021,
  jardat2022diffusion, goswami2023trapped, burov2024,
  santra2024forces}. This situation arises when a tracer in a cold
fluid is coupled to colloids whose fluctuations are enhanced, for
example, by external irradiation~\cite{volpe2011microswimmers}.
\if{The
enhanced fluctuations could also be of internal origin, as is the case
for a bath of bacteria~\cite{wu2000} or
enzymes~\cite{zhao2017enhanced}. When the tracer is much larger than
such particles, their dynamics appear Brownian at the tracer's scale,
but with an effective temperature larger than that of the surrounding
fluid.}\fi
The hot colloids could also model active particles, such as
bacteria~\cite{wu2000,massana2024multiple} or enzymes~\cite{zhao2017enhanced}, if scale
separation allows them to be treated as Brownian on the tracer's
scale. 

While the many-body physics of multi-temperature particle systems has
proven particularly rich~\cite{grosbergjoanny2015, weberfrey2016,
  tanaka2017,Damman2024}, the focus has recently shifted towards
single-tracer dynamics~\cite{ilkerjoanny2021,grosberg2021,
  jardat2022diffusion, goswami2023trapped, burov2024,
  santra2024forces}.
Existing studies have primarily examined free or harmonically trapped tracers, focusing on mean-square displacements and effective temperatures. 
\if{Because such observables fail to capture the irreversible nature of the tracer’s dynamics, a full characterization of its steady state and stochastic properties in general potentials—including entropy production and ratchet currents—remains an open problem.
}\fi
Such observables, however, fail to capture whether the tracer is endowed with an irreversible, nonequilibrium dynamics due to the heat flux between the different reservoirs. Consequently, a full characterization of the tracer dynamics---including its entropy production rate, ratchet current, and steady state in a general potential---remains an outstanding problem.

\if{
The broader literature paints a somewhat confusing picture: Stochastic thermodynamics predicts a steady-state heat flux that should generically drive the tracer out of equilibrium~\cite{visco2006work,Zia_2007,li2019quantifying}, whereas adiabatic perturbation theory predicts a quasistatically slow tracer to follow an effective equilibrium dynamics~\cite{granek2022, solon2022}.}\fi

In this Letter, we show analytically and numerically that the tracer
dynamics are much richer than the existing literature suggests. When
the cold tracer is in contact with a few hot particles at a time, it
displays all standard signatures of active particles, including
accumulation at boundaries~\cite{elgetiWall2013}, steady currents
in ratchet potentials~\cite{di2010bacterial,sokolov2010swimming,Angelani_2011}, and density
rectification by asymmetric obstacles~\cite{galajda2007,
  Tailleur_2009}. Both statics and dynamics are thus driven out of equilibrium. On the contrary, when the tracer
interacts with many bath particles simultaneously, it behaves as a
\textit{bona fide} equilibrium particle: although the full system is driven out of thermal equilibrium by a steady heat flux, nothing distinguishes the cold tracer's dynamics and steady state from those of a Brownian particle at the bath temperature. To show this, we consider an overdamped tracer in a cold fluid---set to zero temperature for simplicity---interacting with a hot bath of $N$ Brownian colloids
at temperature $T$.  In $d$ dimensions, the positions of the tracer, $\br$, and bath
particles, $\{\br_i\}$, evolve via
\begin{subequations}
\label{eq:geneom}
        \begin{align}
    \dot{\br} &= -\mu \nabla_\br V(\br,\{\br_i\}),\\
    \dot{\br}_i &= -\nabla_{\br_i} V(\br,\{\br_i\}) + \sqrt{2 T} \bm{\eta}_i(t)\;,
\end{align}
\end{subequations}
where the $\{\bm{\eta}_i\}$ are independent unit Gaussian white
noises, $\mu$ is the tracer mobility, and the bath mobilities have
been set to unity. Here, $V(\br,\{\br_i\})=U(\br)+\sum_i V_{\rm
  bt}(\br-\br_i)$, where $U$ is an external potential on the tracer
and $V_{\rm bt}$ is a short-ranged repulsive potential coupling the
tracer to the bath particles. Figure~\ref{fig:fig1} reports simulations of Eqs.~\eqref{eq:geneom} in one space dimension, illustrating the transition from active to equilibrium dynamics as the
bath density $\rho=N/L^d$ is increased. 

\if{
These results are illustrated in Fig.~\ref{fig:fig1}, which reports simulations in one space dimension of a zero-temperature overdamped tracer in a bath of $N$ Brownian colloids at temperature $T$.
The positions of the tracer, $\br$, and bath particles, $\{\br_i\}$, evolve via,
\begin{subequations}
\label{eq:geneom}
        \begin{align}
    \dot{\br} &= -\mu \nabla_\br V(\br,\{\br_i\}),\\
    \dot{\br}_i &= -\nabla_{\br_i} V(\br,\{\br_i\}) + \sqrt{2 T} \bm{\eta}_i(t)\;,
\end{align}
\end{subequations}
where the $\{\bm{\eta}_i\}$ are independent unit Gaussian white noises, $\mu$ is the tracer mobility, and the bath mobilities have been set to unity. Here, $V(\br,\{\br_i\})=U(\br)+\sum_i V_{\rm bt}(\br-\br_i)$, where $U$ is an external potential on the tracer and $V_{\rm bt}$ is a short-ranged repulsive potential coupling the tracer to the bath particles. 
}\fi%

To understand these numerical results, we consider a distinct, analytically tractable model where $V_{\rm bt}$ is harmonic and couples all bath particles to the tracer, while 
$U$ remains generic. As we show below, this ``fully-connected" model recapitulates the full phenomenology of Fig.~\ref{fig:fig1}, demonstrating that our results hold for widely different bath dynamics. By eliminating the bath degrees of freedom, we demonstrate analytically the transition to equilibrium in the large-$N$ limit. 
We then develop a perturbation theory for the tracer's steady-state distribution to characterize the active regime. 
We show that non-Boltzmann statistics appear at order $N^{-1}$, but that the entropy-production rate scales only as $N^{-3}$.
\if{implying an intermediate effective equilibrium regime with a non-Boltzmann steady state.}\fi
The tracer thus falls out of its equilibrium  steady state before its dynamics become irreversible, leading to an intermediate regime in which detailed balance is obeyed with respect to a non-Boltzmann steady state.
Finally, we consider a minimal model of a tracer in a hot gel, in which the interactions within the bath have the connectivity of a lattice. We show that the cold tracer never equilibrates and instead drives the entire bath out of equilibrium, causing a long-ranged suppression of bath fluctuations that decays as $r^{-2d}$, where $r$ is the distance to the tracer. 
Together, our results offer a comprehensive understanding of how a cold tracer in a hot bath straddles equilibrium and active dynamics. We close our Letter with a discussion of the experimental implications of our results. Further analytical details, generalizations to other models, and a discussion of finite-temperature tracers are provided in a companion article~\cite{si}. 

\begin{figure}
\raisebox{-0.35em}{
\begin{tikzpicture}[scale=0.99]
\def\h{1.73}
\fill[rounded corners=5pt, blue ,opacity=0.2] (-0.75,-0.6+\h) rectangle (0.75,0.75+\h);
\fill[rounded corners=5pt, brown,opacity=0.2] (-1.57,-.47) rectangle (1.57,0.4);
\node[brown, font=\footnotesize] at (0,-0.7) {Temperature $T$};
\node[blue, font=\footnotesize] at (0,0.9+\h) {$T=0$};
\draw[scale=0.25,domain=-2.23:2.36,smooth,variable=\x,red, shift={(0,\h+4.15)}, line width = 0.7pt] plot ({\x}, {\x*\x*\x*\x/4 - \x*\x/1.5 - \x/4});

\node[red, font=\footnotesize] at (0,0.42+\h) {$U(\br)$};

\foreach \i/\x in {1/-1.2, 2/-0.45, 3/0.3, {\scriptscriptstyle \! N}/1.2} {
    \draw[decorate,decoration={coil,aspect=0.5,amplitude=1.3pt,pre length=0.35cm,post length=0.5cm,segment length=1mm}] (\x,0) -- (0,\h);
    \path (\x-0.55,0) -- (0.1,\h-0.5);
    \draw[fill=white] (\x,0) circle [radius=0.23] node[font=\footnotesize] {$\br_{\i}$} ;
}
  \node[font=\footnotesize] at (0.77,0) {$\dots$};
    \node[font=\footnotesize] at (0.87,0.77) {$k$};
    \node[font=\footnotesize] at (0.32,0.77) {$k$};
    \node[font=\footnotesize] at (-0.42,0.77) {$k$};
    \node[font=\footnotesize] at (-0.89,0.77) {$k$};
\draw[fill=white] (0,\h) circle [radius=0.23] node[font=\footnotesize] {$\br$};

 \draw[decorate, darkgreen, decoration={snake,aspect=0.5,amplitude=1.pt,segment length=1.5mm}, line width=0.6pt] (-.95,0) -- (-.7,0);
 \draw[decorate, darkgreen, decoration={snake,aspect=0.5,amplitude=1.pt,segment length=1.5mm}, line width=0.6pt] (-0.2,0) -- (0.05,0);

\node[darkgreen, font=\scriptsize, anchor=center] at (-0.8,-0.2) {$V_b$};
\node[darkgreen, font=\scriptsize] at (-0.05,-0.2) {$V_b$};

 \node at (-1.42, 0.86+\h) {(a)};
\end{tikzpicture}}
\hfill
\begin{overpic}[scale=0.54]{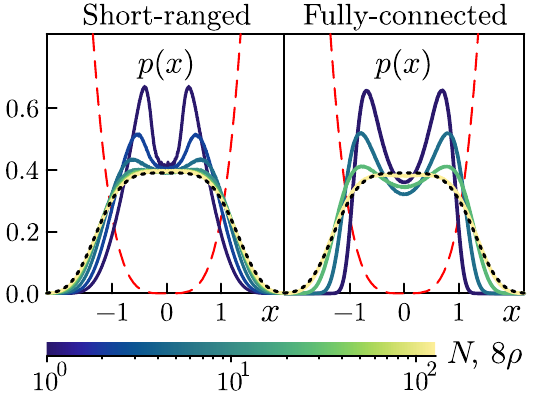} \put(0, 68){(b)}
\end{overpic}
\caption{{\bf (a)} Fully-connected model: The zero-temperature tracer is coupled by springs of stiffness $k$ to $N$ interacting particles. {\bf (b)} Tracer probability density $p(x)$ in a quartic potential $U(x) = x^4/4$ (red dashed line). Left panel is the short-ranged model of Fig.\  \ref{fig:fig1}; right panel is the fully-connected model of Eq.~\eqref{eq:fullyconnectedmodel}. The colored curves are numerical results for different values of $\rho$ in the short-ranged model and $N$ in the fully-connected model. The black dashed lines correspond to the Boltzmann distribution $\propto e^{-U(x)}$.}
\label{fig:mf}
\end{figure}
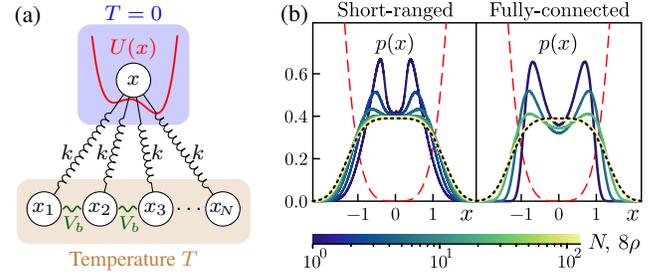

\paragraph{Fully-connected model.} To elucidate analytically the results of Fig.~\ref{fig:fig1}, we introduce the simpler model of Fig.~\ref{fig:mf}(a), where the tracer is linearly coupled to all bath particles. This amounts to Eq.~\eqref{eq:geneom} with
\begin{equation}V(\br, \{\br_i\}) = U(\br) + \sum_{i=1}^N \frac{k}{2}(\br-\br_i)^2 + \sum_{i< j} V_b(\br_i - \br_j)\;,\label{eq:fullyconnectedmodel}
\end{equation}
where, for the sake of generality, we have allowed for pairwise interactions between bath particles. As shown in Fig.~\ref{fig:mf}(b), this fully-connected model exhibits the same phenomenology as the short-ranged model of Fig.~\ref{fig:fig1}: For small $N$, the tracer resembles an active particle, whereas it equilibrates at temperature $T$ when $N\to \infty$. 

To integrate out the bath degrees of freedom, we consider the deviation of the tracer from the bath's center of mass:
\begin{equation} 
\bq \equiv \br - \frac{1}{N}\sum_{i=1}^N \br_i\;.
\end{equation}
The linear coupling between tracer and bath then allows us to turn Eq.~\eqref{eq:geneom} into a closed dynamics for $\br(t)$ and $\bq(t)$:
\begin{subequations}
\label{eq:2vareom}
    \begin{eqnarray}
        \dot{\br} &=& -\mu \nabla U(\br) - \mu N k \bq\;, \\
    \dot{\bq} &=& -\mu \nabla U(\br) - (\mu N + 1)k \bq + \sqrt{\frac{2 T}{N}} \bm{\eta}(t)\;. \label{eq:rdot}
    \end{eqnarray}
\end{subequations}
The tracer dynamics are thus insensitive to the choice of $V_b$. Integrating Eq.~\eqref{eq:rdot} then leads to a non-Markovian dynamics for the tracer position:
\begin{equation} \label{eq:xdotNonMarkov}
    \dot{\br} = \!-\mu \nabla U(\br) + \mu^2 \! N k \!\int_{\!-\infty}^t \!\!\!\!\!\! ds\, e^{-k({\mu N}+1) (t-s)} \nabla U[\br(s)] + \bm{\xi}(t),
\end{equation}
where $\bm{\xi}(t)$ is a persistent Gaussian noise with correlations:
\begin{equation} \label{eq:noisecor}
\langle \xi_\alpha(t_1)\xi_\beta(t_2)\rangle = \frac{\mu^2 N k T}{\mu N + 1} \delta_{\alpha \beta} e^{-k(\mu N+1)|t_1-t_2|}.
\end{equation}
In the absence of an external potential ($U=0$) and with finite $N$, Eqs.~\eqref{eq:xdotNonMarkov} and~\eqref{eq:noisecor} show that the cold tracer is equivalent to a free Active Ornstein-Uhlenbeck Particle~\cite{szamel2014self,sepulveda2013collective,martin2021}. For generic $U(\br)$, this correspondence breaks down and the tracer is endowed with a novel type of active dynamics.

\paragraph{In and out of equilibrium.} 
Let us first use Eqs.~\eqref{eq:xdotNonMarkov} and~\eqref{eq:noisecor} to study the equilibration of the tracer in the large $N$ limit. As $N\rightarrow \infty$, the correlation time of the noise vanishes and $\langle \xi_\alpha(t_1)\xi_\beta(t_2)\rangle \sim~\frac{2T}{N} \delta_{\alpha \beta}\delta(t_1-t_2)$. In this limit, the leading contribution to the integral in Eq.~\eqref{eq:xdotNonMarkov} can be evaluated to yield:
\begin{equation} \label{eq:EquilbDyn}
\dot{\br} = -\frac{\nabla U[\br(t)]}{N} + \sqrt{\frac{2T}{N}} \bm{\eta}(t)\;.
\end{equation}
To leading order in $N^{-1}$, the tracer thus obeys an equilibrium Langevin equation at temperature $T$ with mobility $N^{-1}$. 

We note that the existence of the equilibrium limit can be
  argued on general grounds, beyond the fully-connected harmonic case studied above, provided that a generalized
  Langevin description for the tracer exists. The cold reservoir then contributes
  an $O(1)$ friction and noise, whereas the hot bath contributes
   friction and noise kernels proportional to the number of particles interacting with the tracer, i.e. $O(\rho,N)$.
   In the  $N,\rho\to \infty$ limits, the latter dominate the former, hence enforcing an effective FDT. This phenomenological argument is confirmed numerically in our companion article, where we establish the equilibrium limit both for fully-connected quartic interactions and short-ranged repulsive forces~\cite{si}. Beyond the equilibrium limit, our numerics also confirm that the phenomenology reported in Fig.~\ref{fig:fig1} is generic to all the Brownian baths we considered, irrespective of their precise structure.

To study how the tracer falls out of equilibrium at finite bath densities, we develop a perturbation theory for its steady state at large but finite $N$. For clarity, we set $k=\mu=1$ and refer to~\cite{si} for the general case. From Eq.~\ref{eq:2vareom}, the tracer velocity $\bp \equiv \dot \br$ evolves as
\begin{equation} \label{eq:pdot}
\dot{\bp} = -\nabla U - \left(\bp \cdot \nabla\right) \nabla U - (N+1)\bp + \sqrt{2T N} \bm{\eta}(t)\;.
\end{equation}
The probability density $\Psi(\br,\bp)$ thus obeys $\partial_t \Psi = \mathcal{L} \Psi$, where
\begin{equation}
    \mathcal{L}\!=\!-\bp \! \cdot \! \nabla + \nabla_{\bp} \cdot \left\{\left[\nabla(1 \!+\! \bp \! \cdot \! \nabla) U\right] + (N\!+ \!1)\bp\right\} + T N \nabla_\bp^2,
\end{equation}
and $\nabla_\bp$ denotes a gradient with respect to $\bp$. The steady state is then determined by solving $\cL \Psi = 0$ with $\Psi \propto e^{-\mathcal{H}/T}$, where $\mathcal{H}$ is an effective Hamiltonian that we expand as:
\begin{equation}\label{eq:perttheo}
\mathcal{H}(\br,\bp) = \sum_{n=0}^\infty \frac{ H_n(\br,\bp)}{N^{n}}\;.
\end{equation}
To leading order in $N^{-1}$, we find~\cite{si}:
\begin{equation}
\mathcal{H} \simeq U + \frac{p^2}{2}+\frac{\left(\nabla U\right)^2+2 U-3 T \nabla^2 U+\left(\bp\! \cdot\!\nabla\right)^2 U+p^2}{2N}\;.
\label{eq:h1}
\end{equation}
When $N\to \infty$, the tracer position and velocity follow the equilibrium Maxwell-Boltzmann distribution. For finite $N$, position and velocity become correlated---yet another characteristics of active particles~\cite{szamel2014self,szamel2015glassy,fodor2016, caprini2020, henkes2020dense}. 

\begin{figure}[t]
\vspace{2em}
    \begin{overpic}[scale=0.54, trim={8 14 11 11}, clip]{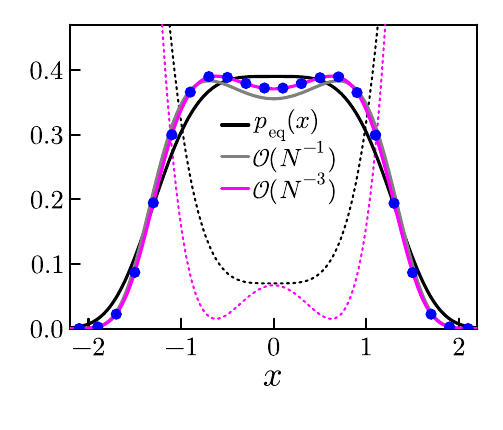} \put(1,76.5){(a)} \put(15,73.5){$p(x)$}
    \put(61.5, 29){{\footnotesize{$U$}}}
    \put(73.5, 20){{\footnotesize \color{magenta}$U_\mathrm{eff}$}}
    \end{overpic}
    \hfill
    \begin{overpic}[scale=0.54, trim={10 18 8 11}, clip]{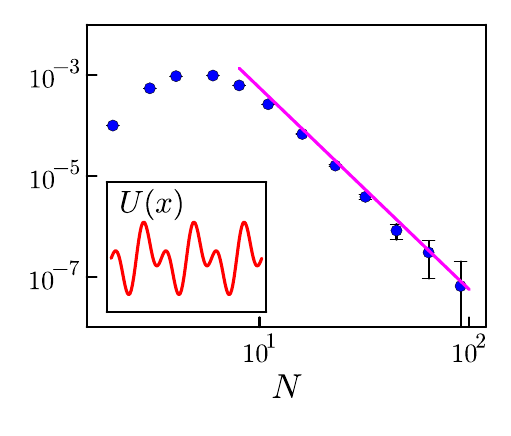} \put(1,74){(b)} 
    \put(16,71.5){$|\langle \dot{x} \rangle|$}
    \put(68, 50){\footnotesize \textcolor{magenta}{Eq.\ \eqref{eq:current}}}
    \end{overpic}

    \begin{tikzpicture}
        \node at (-4.9,0.7) {(c)};
        \draw [-latex, NavyBlue, line width = 1.75pt] (-4,0) -- (2.9,0);
        \node[text=NavyBlue] at (2.5,-0.35) {$N=\infty$};
        \node[text=NavyBlue] at (-0.6,-0.35) {$1 \ll N < \infty$};
        \node[text=NavyBlue] at (-3.7,-0.35) {$N\in \mathcal{O}(1)$};
        \node at (-3.9,0.45) {\parbox{4em}{\centering Active dynamics}};
        \node at (-0.6,0.45) {\parbox{10em}{Reversible, non-Boltzmann}};
        \node at (2.5,0.45) {\parbox{8em}{\centering \small {Effective equilibrium}}};
    \end{tikzpicture}
    \caption{Departure from the large-$N$ equilibrium limit. (a) Stationary probability density $p(x)$ in a potential $U(x) = x^4/4$ (dotted black line). Blue markers are simulation results ($N=10$); solid lines are successive orders in perturbation theory; dotted magenta line is the effective potential at $\mathcal{O}(N^3)$. (b)~Steady-state current, $\langle \dot{x} \rangle$, simulated in a potential $U(x) = \sin(\pi x/2) + \sin(\pi x)$ (inset). The magenta line is Eq.~\eqref{eq:current}, showing the $N^{-4}$ scaling. (c) Successive regimes of tracer dynamics as $N$ is increased.}
    \label{fig:perturbationtheory}
\end{figure}
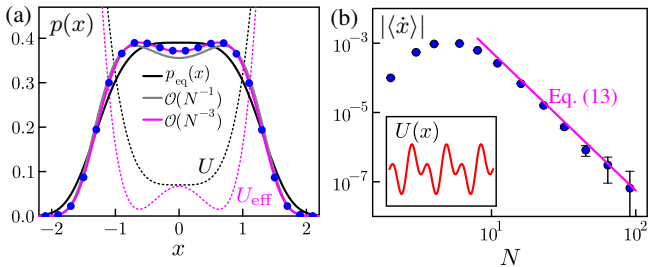

To characterize the deviation from the Boltzmann weight shown in Fig.~\ref{fig:mf}, we compute the distribution of the tracer's position as $P(\br) \propto e^{-U_\mathrm{eff}(\br)/T}$, where
 \begin{equation} \label{eq:Ueff} U_\mathrm{eff}(\br) = U + \frac{1}{N}\left[\frac{\left(\nabla U\right)^2}{2} - T\nabla^2 U + U\right] +  \mathcal{O}\Big(\frac{1}{N^2}\Big).
\end{equation}
In Fig.~\ref{fig:perturbationtheory}(a), we show that the perturbation
theory correctly predicts the accumulation of the tracer away from the
center of a quartic well. This is typical of persistent
motion~\cite{Tailleur_2009,elgetiWall2013} and is already captured at
order $1/N$ by the $-T\nabla^2 U$ contribution to the effective
potential in Eq.~\eqref{eq:Ueff}. Note that the $\mathcal{O}(N^{-3})$
theory~\cite{si} is in excellent agreement with simulations for values
of $N$ as small as $N=10$. 

We now characterize the onset of irreversibility as the systems departs from its large-$N$ equilibrium limit. A hallmark of irreversible dynamics is that asymmetric obstacles act as pumps in nonequilibrium media~\cite{magnasco1993forced,julicher1997modeling,reichhardt2017ratchet,baek2018generic}. In a domain with periodic boundaries, this leads to a steady current, as shown in Fig.~\ref{fig:fig1}(b-c). 
For a potential $U({\bf r}) = U(x)$, asymmetric along $\hat \bx$ and constant along the $d-1$ other dimensions, our perturbation theory predicts a current:\if{, the leading contribution to the current in an asymmetric potential along the $x$-direction $U(\br) = U(x)$ is}\fi
\begin{equation} \label{eq:current}
\langle \dot x \rangle = \frac{L \int_0^L U'(z)^2 U^{(3)}(z) \, dz}{2 N^{4} \int_0^L e^{-\frac{U(y)}{T}} \, dy \int_0^L e^{\frac{U(z)}{T}} \, dz}  + \mathcal{O}\Big(\frac 1 {N^5}\Big)\;.
\end{equation}
Figure~\ref{fig:perturbationtheory}(b) confirms this prediction numerically for  $N\gtrsim 10$. When periodic boundaries are replaced by confining walls, the steady current is replaced by a long-ranged density build-up on one side of the obstacle~\cite{galajda2007}, as shown in Fig~\ref{fig:fig1}(d). This can be quantified as the difference in probability density at two points $x_\ell$ and $x_r$ on either side of the obstacle:
\begin{equation}
\frac{P(x_r) - P(x_\ell)}{P(x_\ell)} =   \int_{x_\ell}^{x_r} dz \frac{U'(z)^2 U^{(3)}(z)}{2T N^3} +\mathcal{O}\Big(\frac 1 {N^4}\Big).
\end{equation}
Note that, as $N$ increases, the density rectification and ratchet current vanish faster than the deviation from the Boltzmann weight~\footnote{The ratchet current appears at a higher order than the rectification because the effective tracer mobility in Eq.~\eqref{eq:EquilbDyn} scales as $1/N$ for $N\gg 1$.}.
This suggests the existence of an intermediate regime where the system has a non-Boltzmann steady-state but a reversible dynamics (Fig. \ref{fig:perturbationtheory}c). To test this, we compute the entropy production rate of the tracer, defined as~\cite{maes1999,seifert2005}:
\begin{equation}\sigma \equiv \lim_{t\rightarrow \infty} \frac{1}{t} \left\langle\log \frac{\mathcal{P}\left[\{\br(s\leq t)\}\right]} 
{\mathcal{P}\left[\{\br^\mathrm{R}(s\leq t)\}\right]}\right\rangle,
\end{equation}
where $\mathcal{P}$ is the path probability density, and $\br^\mathrm{R}(s) \equiv \br(t-s)$. Using standard methods~\cite{si}, we find 
\begin{equation} \label{eq:sigma}
\sigma = \frac{\big\langle \!\left(\bp \cdot \nabla\right)^3 U \big\rangle }{2TN}
\simeq  \frac{T}{2 N^3} \left \langle \left(\nabla^3 U\right)^2\right \rangle_{\mathrm{eq}} + \mathcal{O}\Big(\frac{1}{N^4}\Big)\;,
\end{equation}
where the final equality follows from the perturbation theory and $\langle \cdot \rangle_{\mathrm{eq}}$ denotes an average with respect to $P_{\rm eq} \propto e^{-U(x)/T}$. The existence of a non-Boltzmann yet time-reversible regime is confirmed by the fact that $\sigma$ vanishes as $N^{-3}$ while deviations from the Boltzmann weight scale as $N^{-1}$. 

In our companion article~\cite{si}, we show numerically that the $N^{-1}$ scaling for the departure from the Boltzmann weight and the $N^{-4}$ scaling for the current hold when the tracer and bath particles interact via a quartic potential. They also extend to systems where tracer and bath particles interact via a short-range potential. These scaling thus appear to be universal properties of Brownian baths. 
 
\begin{figure}
    \centering
\begin{tikzpicture}[scale=0.87]
\def\h{1.7}

\fill[brown,opacity=0.2] 
    plot [smooth cycle] coordinates {(.7,-0.5) (1.05,-0.3) (1.35,0.1) (1.4,0.9)  (1.1,1.60) (.7,1.58) (.4,1.07) (-.4,1.07) (-.7,1.58) (-1.1,1.60) (-1.4,0.9) (-1.35,0.1)  (-1.05,-0.3) (-.7,-0.5) (0,-0.6)};

\fill[blue,opacity=0.2] 
    plot [smooth cycle] coordinates {(0.7, 2.3)  (-0.7, 2.3) (-0.4, 1.2) (0.4, 1.2)};
    
\node[brown, font=\footnotesize] at (0,0.5) {$T>0$};
\node[blue, font=\footnotesize] at (0,0.85+\h) {$T=0$};
\draw[scale=0.25,domain=-2.23:2.36,smooth,variable=\x,red, shift={(-0.2,\h+3.9)}, line width = 0.7pt] plot ({\x}, {\x*\x*\x*\x/4 - \x*\x/1.5 - \x/4});

\node[red, font=\footnotesize] at (0,0.45+\h) {$U(x)$};

\def\xOne{0.9}
\def\yOne{1.3}
\def\xTwo{1.1}
\def\yTwo{0.6}
\def\xThree{0.75}
\def\yThree{-0.1}
\def\yZero{\h}

\foreach \xstart/\ystart/\xend/\yend in {0/\h/\xOne/\yOne, \xOne/\yOne/\xTwo/\yTwo, \xTwo/\yTwo/\xThree/\yThree, 0/\h/-\xOne/\yOne, -\xOne/\yOne/-\xTwo/\yTwo, -\xTwo/\yTwo/-\xThree/\yThree} {
    \draw[decorate,decoration={coil,aspect=0.5,amplitude=1.1pt,pre length=0.22cm,post length=0.15cm,segment length=0.9mm}] (\xstart,\ystart) -- (\xend,\yend);
}

\draw[decorate, decoration={coil,aspect=0.5,amplitude=1.3pt,pre length=0.25cm,post length=.02cm,segment length=0.9mm}] (\xThree,\yThree) -- (.32,\yThree-0.3);
\draw[decorate, decoration={coil,aspect=0.5,amplitude=1.3pt,pre length=0.25cm,post length=.02cm,segment length=0.9mm}] (-\xThree,\yThree) -- (-0.32,\yThree-0.3);

\node[font=\small] at (0, \yThree-0.35) {$\dots$};

\foreach \i/\x/\y in {1/\xOne/\yOne, 2/\xTwo/\yTwo, 3/\xThree/\yThree, {\scriptscriptstyle \!N}/-\xOne/\yOne, \scalebox{0.78}{$\scriptscriptstyle \mkern-4.5mu N\mkern-1.3mu \text{--}\mkern-1.2mu 1$}/-\xTwo/\yTwo, \scalebox{0.78}{$\scriptscriptstyle \mkern-4.5mu N\mkern-1.3mu \text{--}\mkern-1.2mu 2$}/-\xThree/\yThree} {
    \draw[fill=white] (\x,\y) circle [radius=0.25] node[font=\scriptsize] {$x_{\i}$} ;
}
\draw[fill=white] (0,\yZero) circle [radius=0.25] node[font=\footnotesize] {$x$};
\node at (-1.1,\h+0.54) {(a)};
\end{tikzpicture}
\raisebox{0.5em}{\begin{tikzpicture}[scale=0.82]
\def\latspace{0.78}
\def\edgeopac{0.5}

\begin{scope}
\clip[rounded corners=20pt] (-1.42,-1.42) rectangle (1.42,1.42);
    \fill[rounded corners=10pt, brown,opacity=0.2] (-2,-2) rectangle (2,2);
    \fill[rounded corners=10pt, white] (-0.52,-0.52) rectangle (0.52,0.52);
    \fill[rounded corners=10pt, blue ,opacity=0.2] (-0.45,-0.45) rectangle (0.45,0.45);

\draw[decorate, decoration={coil,aspect=0.5,amplitude=1.1pt,pre length=0.1cm,post length=0.1cm,segment length=0.9mm}] (0,0) -- (\latspace,0);
\draw[decorate, decoration={coil,aspect=0.5,amplitude=1.1pt,pre length=0.1cm,post length=0.1cm,segment length=0.9mm}] (-\latspace,0) -- (0,0);
\draw[decorate, decoration={coil,aspect=0.5,amplitude=1.1pt,pre length=0.1cm,post length=0.1cm,segment length=0.9mm}] (0,0) -- (0,\latspace);
\draw[decorate, decoration={coil,aspect=0.5,amplitude=1.1pt,pre length=0.1cm,post length=0.1cm,segment length=0.9mm}] (0,-\latspace) -- (0,0);

\draw[decorate, decoration={coil,aspect=0.5,amplitude=1.1pt,pre length=0.1cm,post length=0.1cm,segment length=0.9mm}] (\latspace,0) -- (\latspace,\latspace);
\draw[decorate, decoration={coil,aspect=0.5,amplitude=1.1pt,pre length=0.1cm,post length=0.1cm,segment length=0.9mm}] (-\latspace,-\latspace) -- (-\latspace,0);
\draw[decorate, decoration={coil,aspect=0.5,amplitude=1.1pt,pre length=0.1cm,post length=0.1cm,segment length=0.9mm}] (0,\latspace) -- (\latspace,\latspace);
\draw[decorate, decoration={coil,aspect=0.5,amplitude=1.1pt,pre length=0.1cm,post length=0.1cm,segment length=0.9mm}] (-\latspace,-\latspace) -- (0,-\latspace);
\draw[decorate, decoration={coil,aspect=0.5,amplitude=1.1pt,pre length=0.1cm,post length=0.1cm,segment length=0.9mm}] (-\latspace,0) -- (-\latspace,\latspace);
\draw[decorate, decoration={coil,aspect=0.5,amplitude=1.1pt,pre length=0.1cm,post length=0.1cm,segment length=0.9mm}] (-\latspace,\latspace) -- (0,\latspace);
\draw[decorate, decoration={coil,aspect=0.5,amplitude=1.1pt,pre length=0.1cm,post length=0.1cm,segment length=0.9mm}] (0,-\latspace) -- (\latspace,-\latspace);
\draw[decorate, decoration={coil,aspect=0.5,amplitude=1.1pt,pre length=0.1cm,post length=0.1cm,segment length=0.9mm}] (\latspace,-\latspace) -- (\latspace,0);

\draw[decorate, decoration={coil,aspect=0.5,amplitude=1.1pt,pre length=0.0cm,post length=0.0cm,segment length=0.9mm}] (-2*\latspace+0.25,\latspace) -- (-\latspace,\latspace);
\draw[decorate, decoration={coil,aspect=0.5,amplitude=1.1pt,pre length=0.0cm,post length=0.0cm,segment length=0.9mm}] (-2*\latspace+0.25,0) -- (-\latspace,0);
\draw[decorate, decoration={coil,aspect=0.5,amplitude=1.1pt,pre length=0.0cm,post length=0.0cm,segment length=0.9mm}] (-2*\latspace+0.25,-\latspace) -- (-\latspace,-\latspace);

\draw[decorate, decoration={coil,aspect=0.5,amplitude=1.1pt,post length=0.0cm,pre length=0.0cm,segment length=0.9mm}] (\latspace,\latspace) -- (2*\latspace-0.25,\latspace);
\draw[decorate, decoration={coil,aspect=0.5,amplitude=1.1pt,post length=0.0cm,pre length=0.0cm,segment length=0.9mm}] (\latspace,0) -- (2*\latspace-0.25,0);
\draw[decorate, decoration={coil,aspect=0.5,amplitude=1.1pt,post length=0.0cm,pre length=0.0cm,segment length=0.9mm}] (\latspace,-\latspace) -- (2*\latspace-0.25,-\latspace) ;

\draw[decorate, decoration={coil,aspect=0.5,amplitude=1.1pt,post length=0cm,pre length=0cm,segment length=0.9mm}] (\latspace,\latspace) -- (\latspace, 2*\latspace-0.25);
\draw[decorate, decoration={coil,aspect=0.5,amplitude=1.1pt,post length=0cm,pre length=0cm,segment length=0.9mm}] (0,\latspace) -- (0, 2*\latspace-0.25);
\draw[decorate, decoration={coil,aspect=0.5,amplitude=1.1pt,post length=0cm,pre length=0cm,segment length=0.9mm}] (-\latspace,\latspace) -- (-\latspace, 2*\latspace-0.25) ;

\draw[decorate, decoration={coil,aspect=0.5,amplitude=1.1pt,pre length=0cm,post length=0cm,segment length=0.9mm}]  (\latspace, -2*\latspace+0.25) -- (\latspace,-\latspace);
\draw[decorate, decoration={coil,aspect=0.5,amplitude=1.1pt,pre length=0cm,post length=0cm,segment length=0.9mm}] (0, -2*\latspace+0.25)  --(0,-\latspace);
\draw[decorate, decoration={coil,aspect=0.5,amplitude=1.1pt,pre length=0cm,post length=0cm,segment length=0.9mm}] (-\latspace, -2*\latspace+0.25) -- (-\latspace,-\latspace);

\draw[fill=white] (0,0) circle [radius=0.25] node[font=\footnotesize] {$\br$} ;
\draw[fill=white,opacity=\edgeopac] (-2*\latspace,0) circle [radius=0.22];
\draw[fill=white] (-\latspace,0) circle [radius=0.22];
\draw[fill=white] (\latspace,0) circle [radius=0.22];
\draw[fill=white,opacity=\edgeopac] (2*\latspace,0) circle [radius=0.22];

\draw[fill=white,opacity=\edgeopac] (-2*\latspace,2*\latspace) circle [radius=0.22];
\draw[fill=white,opacity=\edgeopac] (-\latspace,2*\latspace) circle [radius=0.22];
\draw[fill=white,opacity=\edgeopac] (0,2*\latspace) circle [radius=0.22];
\draw[fill=white,opacity=\edgeopac] (\latspace,2*\latspace) circle [radius=0.22];
\draw[fill=white,opacity=\edgeopac] (2*\latspace,2*\latspace) circle [radius=0.22];

\draw[fill=white,opacity=\edgeopac] (-2*\latspace,\latspace) circle [radius=0.22];
\draw[fill=white] (-\latspace,\latspace) circle [radius=0.22];
\draw[fill=white] (0,\latspace) circle [radius=0.22];
\draw[fill=white] (\latspace,\latspace) circle [radius=0.22];
\draw[fill=white,opacity=\edgeopac] (2*\latspace,\latspace) circle [radius=0.22];

\draw[fill=white,opacity=\edgeopac] (-2*\latspace,-\latspace) circle [radius=0.22];
\draw[fill=white] (-\latspace,-\latspace) circle [radius=0.22];
\draw[fill=white] (0,-\latspace) circle [radius=0.22];
\draw[fill=white] (\latspace,-\latspace) circle [radius=0.22];
\draw[fill=white,opacity=\edgeopac] (2*\latspace,-\latspace) circle [radius=0.22];

\draw[fill=white,opacity=\edgeopac] (-2*\latspace,-2*\latspace) circle [radius=0.22];
\draw[fill=white,opacity=\edgeopac] (-\latspace,-2*\latspace) circle [radius=0.22];
\draw[fill=white,opacity=\edgeopac] (0,-2*\latspace) circle [radius=0.22];
\draw[fill=white,opacity=\edgeopac] (\latspace,-2*\latspace) circle [radius=0.22];
\draw[fill=white,opacity=\edgeopac] (2*\latspace,-2*\latspace) circle [radius=0.22];
\end{scope}
\node at (-1.35,1.38) {(b)};
\end{tikzpicture}}\hfill
\raisebox{-1em}{\begin{overpic}[scale=0.54, trim={0.2cm, 0.83cm, 0cm, 0cm}, clip]{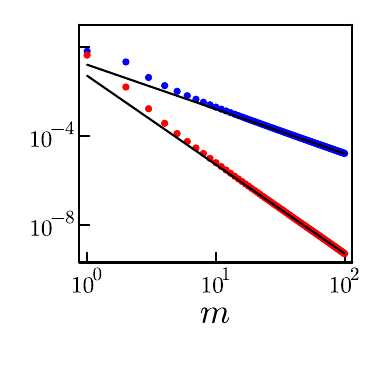} \put(5, 78){(c)}
\put(69, 58){\color{blue}{\footnotesize$d=1$}}
\put(46, 30){\color{red}{\footnotesize$d=2$}}
\put(62.5, 74){\footnotesize $\Delta T(m)$}
\end{overpic}}

    \caption{(a) Loop model: A zero-temperature particle is inserted within a loop of particles at temperature $T$ and subjected to a potential $U$. (b) Generalization to higher-dimensional lattices. (c) Effective temperature suppression of a particle $m$ sites away from the tracer. Blue markers are Eq. \eqref{eq:1dTeff} $(n=d=1)$. Red markers are exact numerical results for $N=400^2$ particles with $n=1, d=2$. Black lines correspond to the field-theoretic prediction Eq. \eqref{eq:tefffieldtheory}. $T=k=1$.}
    \label{fig:fkm}
\end{figure}
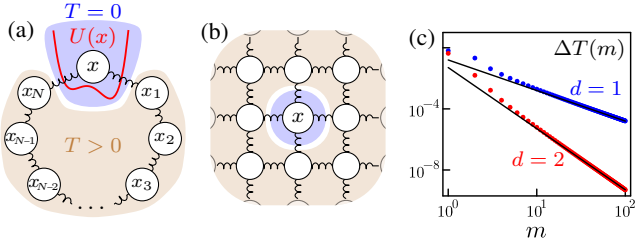

\paragraph{Finite tracer connectivity: a cold tracer in a hot gel.} The models studied so far aim at characterizing a tracer immersed in a fluid, where the large-$N$ (or large-tracer) limit can be realized. Another interesting situation is that of a cold inclusion in an active gel, where the connectivity between tracer and bath particles are fixed and finite. The temperature difference between the tracer and the gel can then be thought to represent active fluctuations in a cytoskeletal network~\cite{juelicher2007active,mizuno2007nonequilibrium,wilhelm2008out,fodor2014energetics,ben2015modeling,bohec2019distribution} or a soft active solid~\cite{baconnier2022selective,massana2024multiple}. To address this case, we consider the ``loop" topology of Fig.~\ref{fig:fkm}(a), inspired by the work of Ford, Kac, and Mazur on Hamiltonian systems~\cite{fkm1965}, as well as its generalization to higher-coordination lattices as in Fig.~\ref{fig:fkm}(b). 

We first note that the dynamics of a cold tracer interacting with an arbitrary linearly-coupled bath can be mapped onto a generalized Langevin equation of the form~\cite{si}:
\begin{equation}\label{eq:genlan}
\int_{-\infty}^t ds \, K(t-s) \dot{\br}(s) = -\nabla U(\br) + \bm{ \xi}(t)\;.
\end{equation}
In the large-$N$ limit of the $1d$ loop model of Fig.~\ref{fig:fkm}(a), we find:
\begin{subequations}
\begin{align}
    K(s) = \frac{2}{\mu} \delta(s) + 2 e^{-2k|s|} \left[I_0(2k|s|)\!+\!I_1(2k|s|)\right]\;,\\
    \langle \zeta(t) \zeta(t\!+\!s)\rangle = 2 T  e^{-2k|s|} \left[I_0(2k|s|)\!+\!I_1(2k|s|)\right]\;,
\end{align}  
\end{subequations}
where $I_n(z)$ is the $n$th modified Bessel function of the first kind, $k$ is the spring stiffness, and $\mu$ the tracer mobility. Since  $\langle \zeta(t) \zeta(t+s) \rangle \neq T K(s)$, the kernels violate the FDT~\cite{zwanzig2001} and the hot gel always drives the tracer out of thermal equilibrium.

We now turn to the converse question of how the cold tracer impacts the hot particles.
Due to heat exchange with the reservoirs, energy is not locally conserved. We thus cannot invoke the heat equation to predict the influence of a localized temperature sink on the gel fluctuations~\cite{bonetto2000fourier}. Nevertheless, this question can be addressed for the loop model when $U(x)$ is quadratic, so that all $\{x_i\}$ become Gaussian distributed. In absence of the cold tracer, the fluctuations of the $m$th spring follow $k\langle(x_{m+1} - x_m)^2\rangle_\mathrm{eq}=T$. To quantify the impact of the tracer at $m = 0$ on the gel fluctuations, we assign to the $m$th bath particle an effective temperature through
\begin{equation}T_\mathrm{eff}(m) \equiv \frac{1}{2}k \left\langle(x_{m} - x_{m-1})^2 + (x_{m+1} - x_m)^2\right\rangle, \label{eq:TeffDef} \end{equation}
and compute $\Delta T(m) \equiv T-T_{\rm eff}(m)$. As $N\rightarrow \infty$, we find this to approach the exact form \cite{si}
\begin{equation}
\Delta T(0) = \frac{\pi-2}{\pi}\,T\,, \quad \Delta T(m\geq  1) = \frac{2 T}{\pi} \frac{1}{4m^2-1} \;.
\label{eq:1dTeff}\end{equation}
\if{and $T_\mathrm{eff}(0) = 2T/\pi$.}\fi Notably, the  cold tracer induces a long-ranged damping of fluctuations that decays as $m^{-2}$ for $m \gg 1$. The entire gel is thus cooled down by the tracer.

To generalize the above to $n$ space dimensions and higher-coordination lattices, we consider $N=L^d$ bath particles connected by springs into a $d$-dimensional hypercubic lattice, as in Fig.~\ref{fig:fkm}(b). \if{For generality, we allow $d \neq n$.}\fi Bulk materials correspond to $d=n$, while $d<n$ and $d>n$ respectively correspond to lattices embedded in higher dimensions or confined to lower dimensions. We denote by $\vec{r}_\mathbf{m}\in \mathbb{R}^n$ the displacement of the particle at lattice coordinate $\mathbf{m} =(m_1, \dots, m_d)$, with $\mathbf{m} = \mathbf{0}$ corresponding to the cold tracer. We focus here on the case in which the $\{\vec{r}_\mathbf{m}\}$ are coupled by harmonic potentials of stiffness $k$, which can be coarse-grained exactly, but our results extend to a more general elastic theory~\cite{si}.
The coarse-graining follows by defining a continuous lattice index $\br \equiv \mathbf{m}/L$, a displacement field $\vec{u}(\mathbf{r}) \equiv L^{\frac{d-2}{2}} \vec{r}_{\mathbf{m}}$, and a rescaled time coordinate $t\rightarrow t/L^2$. Setting $\mu = 1$ for brevity and sending $L\to \infty$ , we find $\vec{u}$ to evolve as
\begin{equation}\partial_t \vec{u}(\br, t) = k\nabla^2 \vec{u} + \vec{\Lambda}(\br, t), \label{eq:fieldtheory}
\end{equation}
where $\vec{\Lambda}(\br, t)$ is a Gaussian white noise field with correlations:
\begin{equation} 
\langle \Lambda_i(\br, t) \Lambda_j(\br', t') \rangle = 2T \delta_{ij} \delta(t-t') \delta(\br - \br')\left[1 - L^{-d} \delta(\br)\right].
\end{equation}
It can be shown by solving Eq.~\eqref{eq:fieldtheory} in Fourier space that
\begin{align}\langle \nabla u_i(\br,t) \cdot \nabla u_j&(\br',t)\rangle \nonumber \\= \frac{T}{k}\delta_{ij} &\left[\delta(\br - \br') - \frac{2 d!}{\pi^d L^d} \frac{\br \cdot \br'}{(r^2+r'^2)^{1+d}}\right],\label{eq:gradflucs}\end{align}
which is valid away from $\br = \br' = 0$. An effective temperature for the particle at $\mathbf{m}$ can be defined by extending the average in Eq.~\eqref{eq:TeffDef} to include the $2d$ nearest-neighbors of $\mathbf{m}$. Rewriting Eq. \eqref{eq:gradflucs} in terms of the microscopic coordinates, we find the temperature suppression to decay as
\begin{equation}
    \Delta T(\mathbf{m}) = \frac{d!}{(2\pi)^d}\frac{T}{m^{2d}}. \label{eq:tefffieldtheory}
\end{equation}
Setting $n = d = 1$ recovers the exact result of Eq. \eqref{eq:1dTeff} for large $m$. In Fig.~\ref{fig:fkm}(c), we verify the $n=1, d=2$ prediction against numerical results for $N = 400^2$ lattice particles. We conclude that, for a $d$-dimensional lattice, the effect of the cold tracer is long-ranged and decays as $m^{-2d}$. We note that this  exponent differs from the $2-d$ exponent associated with the Green's function of the Laplacian in Eq.~\eqref{eq:fieldtheory}. It is also not a solution to the heat equation since $\nabla^2 T(\mathbf{m})\neq 0$. We finally stress that it is nonequilibrium in nature, since applying a localized potential to overstretch an equilibrium lattice has only a short-ranged effect~\cite{si}.

\paragraph{Conclusion.} 
In this Letter, we have developed a theoretical framework for the dynamics of a zero-temperature tracer in a bath of Brownian particles. This framework first reveals an effective equilibrium regime for the tracer at large bath density. Then, it shows how the tracer's steady state first departs its Boltzmann limit as the bath density decreases, before its dynamics becomes fully active and irreversible. 

A possible experimental application of our results concerns
``active enzymes", such as urease, which are driven out of equilibrium
by a chemical potential difference between their substrates and
products~\cite{ghosh2021}. Such particles, ubiquitous in the cellular
environment, have been shown to exhibit enhanced Brownian
fluctuations~\cite{muddana2010substrate, sengupta2013enzyme} and to
enhance the diffusivity of nearby passive
colloids~\cite{dey2015micromotors, zhao2017enhanced}. Our results
predict that a passive tracer in a solution of such enzymes---either
in vitro or in the cytosol---will rectify their fluctuations into a
persistent motion, displaying a rich non-equilibrium dynamics.

When the environment is gel-like, so that bath particles have the
connectivity of a lattice, we found that the tracer drives the entire
bath out of equilibrium through a long-ranged damping of
fluctuations. Experimentally, our results could be tested in soft active solids, where the persistence of the active
components can be controlled by their
shapes~\cite{baconnier2022selective}, or in colloid crystals interacting with light-controlled bacteria~\cite{massana2024multiple}. They could also prove relevant
for active gels~\cite{mizuno2007nonequilibrium,wilhelm2008out}, where
inserting passive probes into the cytoskeletal network could lead to
long-ranged damping of the active fluctuations.

Finally, we have neglected the role of hydrodynamic interactions in
our study and it would be interesting to know how the presence of a
momentum-conserving fluid might impact the properties of the tracer
and of the bath~\cite{arnoulx2024anomalous}. Similarly, in the
presence of a short-range potential between the tracer and the bath,
the bath density field is a hydrodynamic mode. This should endow the
tracer dynamics with a long-range memory~\cite{granek2022} whose
significance remains to be determined.

\if{
  Beyond these applications, our findings have implications for the many-body physics of multi-temperature systems: it has been shown~\cite{grosbergjoanny2015,weberfrey2016,tanaka2017} that mixtures of hot and cold particles spontaneously demix. In light of our results, it would be interesting to investigate whether this demixing could be interpreted as a motility-induced phase separation of the cold particles~\cite{mipsrev2015}, which are turned active by their interactions with their hot neighbors.
  }\fi

\noindent\textbf{\textit{Acknowledgements.}}
We thank Hugues Chaté, Omer Granek, Yariv Kafri, Mehran Kardar, Joel Lebowitz, and Alex Solon for fruitful discussions. We thank Jessica Metzger and Julia Yeomans for discussions on the many-body implications of our results. AA acknowledges the financial support of the MathWorks fellowship. AA and JT acknowledge the hospitality of MSC laboratory and the financial support of an MIT MISTI GSF grant.

\bibliography{./bibliography}

@article{massana2024multiple,
  title={Multiple temperatures and melting of a colloidal active crystal},
  author={Massana-Cid, Helena and Maggi, Claudio and Gnan, Nicoletta and Frangipane, Giacomo and Di Leonardo, Roberto},
  journal={Nature Communications},
  volume={15},
  number={1},
  pages={6574},
  year={2024},
  publisher={Nature Publishing Group UK London}
}

@article{maggi2014generalized,
  title={Generalized energy equipartition in harmonic oscillators driven by active baths},
  author={Maggi, Claudio and Paoluzzi, Matteo and Pellicciotta, Nicola and Lepore, Alessia and Angelani, Luca and Di Leonardo, Roberto},
  journal={Physical review letters},
  volume={113},
  number={23},
  pages={238303},
  year={2014},
  publisher={APS}
}

@article{Damman2024,
  title = {Algebraic Depletion Interactions in Two-Temperature Mixtures},
  author = {Damman, Pascal and D\'emery, Vincent and Palumbo, Guillaume and Thomas, Quentin},
  journal = {Phys. Rev. Lett.},
  volume = {133},
  issue = {26},
  pages = {267103},
  numpages = {6},
  year = {2024},
  month = {Dec},
  publisher = {American Physical Society},
  doi = {10.1103/PhysRevLett.133.267103},
  url = {https://link.aps.org/doi/10.1103/PhysRevLett.133.267103}
}

@article{burov2024,
  title={Emergence of directed motion in a crowded suspension of overdamped particles with different effective temperatures},
  author={Schwarcz, Deborah and Burov, Stanislav},
  journal={Physical Review Research},
  volume={6},
  number={1},
  pages={013156},
  year={2024},
  publisher={APS}
}

@article{henkes2020dense,
  title={Dense active matter model of motion patterns in confluent cell monolayers},
  author={Henkes, Silke and Kostanjevec, Kaja and Collinson, J Martin and Sknepnek, Rastko and Bertin, Eric},
  journal={Nature communications},
  volume={11},
  number={1},
  pages={1405},
  year={2020},
  publisher={Nature Publishing Group UK London}
}

@article{szamel2015glassy,
  title={Glassy dynamics of athermal self-propelled particles: Computer simulations and a nonequilibrium microscopic theory},
  author={Szamel, Grzegorz and Flenner, Elijah and Berthier, Ludovic},
  journal={Physical Review E},
  volume={91},
  number={6},
  pages={062304},
  year={2015},
  publisher={APS}
}

@article{grosberg2021,
  title={Tethered tracer in a mixture of hot and cold Brownian particles: can activity pacify fluctuations?},
  author={Wang, Michael and Zinga, Ketsia and Zidovska, Alexandra and Grosberg, Alexander Y},
  journal={Soft Matter},
  volume={17},
  number={41},
  pages={9528--9539},
  year={2021},
  publisher={Royal Society of Chemistry}
}

@article{Argun2016,
author = {Argun, Aykut and Moradi, Ali-Reza and Pin{\c{c}}e, Er{\c{c}}aǧ and Bagci, Gokhan Baris and Imparato, Alberto and Volpe, Giovanni},
doi = {10.1103/PhysRevE.94.062150},
issn = {2470-0045},
journal = {Phys. Rev. E},
month = {dec},
number = {6},
pages = {062150},
title = {{Non-Boltzmann stationary distributions and nonequilibrium relations in active baths}},
url = {https://link.aps.org/doi/10.1103/PhysRevE.94.062150},
volume = {94},
year = {2016}
}

@Article{Chaki2018,
  author    = {Chaki, Subhasish and Chakrabarti, Rajarshi},
  journal   = {Phys. A Stat. Mech. its Appl.},
  title     = {{Entropy production and work fluctuation relations for a single particle in active bath}},
  year      = {2018},
  issn      = {03784371},
  month     = {dec},
  pages     = {302--315},
  volume    = {511},
  doi       = {10.1016/j.physa.2018.07.055},
  publisher = {North-Holland},
  url       = {https://linkinghub.elsevier.com/retrieve/pii/S0378437118309270},
}

@Article{Knezevic2020,
  author   = {Kne{\v{z}}evi{\'{c}}, Miloš and Stark, Holger},
  journal  = {New J. Phys.},
  title    = {{Effective Langevin equations for a polar tracer in an active bath}},
  year     = {2020},
  issn     = {1367-2630},
  month    = {11},
  number   = {11},
  pages    = {113025},
  volume   = {22},
  arxivid  = {2008.03207},
  doi      = {10.1088/1367-2630/abc91e},
  url      = {https://doi.org/10.1088/1367-2630/abc91e https://iopscience.iop.org/article/10.1088/1367-2630/abc91e},
}

@article{Loi2008,
    title = {{Effective temperature of active matter}},
    year = {2008},
    journal = {Phys. Rev. E},
    author = {Loi, Davide and Mossa, Stefano and Cugliandolo, Leticia F.},
    number = {5},
    month = {5},
    pages = {051111},
    volume = {77},
    url = {https://link.aps.org/doi/10.1103/PhysRevE.77.051111},
    doi = {10.1103/PhysRevE.77.051111},
    issn = {1539-3755}
}

@Article{Underhill2008,
  author   = {Underhill, Patrick T. and Hernandez-Ortiz, Juan P. and Graham, Michael D.},
  journal  = {Phys. Rev. Lett.},
  title    = {{Diffusion and Spatial Correlations in Suspensions of Swimming Particles}},
  year     = {2008},
  issn     = {0031-9007},
  month    = {6},
  number   = {24},
  pages    = {248101},
  volume   = {100},
  arxivid  = {0805.3784},
  doi      = {10.1103/PhysRevLett.100.248101},
  url      = {https://link.aps.org/doi/10.1103/PhysRevLett.100.248101},
}

@article{fodor2016,
  title = {How Far from Equilibrium Is Active Matter?},
  author = {Fodor, \'Etienne and Nardini, Cesare and Cates, Michael E. and Tailleur, Julien and Visco, Paolo and van Wijland, Fr\'ed\'eric},
  journal = {Phys. Rev. Lett.},
  volume = {117},
  issue = {3},
  pages = {038103},
  numpages = {6},
  year = {2016},
  month = {Jul},
  publisher = {American Physical Society},
  doi = {10.1103/PhysRevLett.117.038103},
  url = {https://link.aps.org/doi/10.1103/PhysRevLett.117.038103}
}

@article{visco2006work,
  title={Work fluctuations for a Brownian particle between two thermostats},
  author={Visco, Paolo},
  journal={Journal of Statistical Mechanics: Theory and Experiment},
  volume={2006},
  number={06},
  pages={P06006},
  year={2006},
  publisher={IOP Publishing}
}

@article{martin2021,
  title = {Statistical mechanics of active Ornstein-Uhlenbeck particles},
  author = {Martin, David and O'Byrne, J\'er\'emy and Cates, Michael E. and Fodor, \'Etienne and Nardini, Cesare and Tailleur, Julien and van Wijland, Fr\'ed\'eric},
  journal = {Phys. Rev. E},
  volume = {103},
  issue = {3},
  pages = {032607},
  numpages = {25},
  year = {2021},
  month = {Mar},
  publisher = {American Physical Society},
  doi = {10.1103/PhysRevE.103.032607},
  url = {https://link.aps.org/doi/10.1103/PhysRevE.103.032607}
}

@article{granek2022,
  title = {Anomalous Transport of Tracers in Active Baths},
  author = {Granek, Omer and Kafri, Yariv and Tailleur, Julien},
  journal = {Phys. Rev. Lett.},
  volume = {129},
  issue = {3},
  pages = {038001},
  numpages = {7},
  year = {2022},
  month = {Jul},
  publisher = {American Physical Society},
  doi = {10.1103/PhysRevLett.129.038001},
  url = {https://link.aps.org/doi/10.1103/PhysRevLett.129.038001}
}

@article{mizuno2007nonequilibrium,
  title={Nonequilibrium mechanics of active cytoskeletal networks},
  author={Mizuno, Daisuke and Tardin, Catherine and Schmidt, Christoph F and MacKintosh, Frederik C},
  journal={Science},
  volume={315},
  number={5810},
  pages={370--373},
  year={2007},
  publisher={American Association for the Advancement of Science}
}

@article{arnoulx2024anomalous,
  title={Anomalous Long-Ranged Influence of an Inclusion in Momentum-Conserving Active Fluids},
  author={Arnoulx de Pirey, Thibaut and Kafri, Yariv and Ramaswamy, Sriram},
  journal={Physical Review X},
  volume={14},
  number={4},
  pages={041034},
  year={2024},
  publisher={APS}
}

@article{wilhelm2008out,
  title={Out-of-equilibrium microrheology inside living cells},
  author={Wilhelm, Claire},
  journal={Physical review letters},
  volume={101},
  number={2},
  pages={028101},
  year={2008},
  publisher={APS}
}

@article{baconnier2022selective,
  title={Selective and collective actuation in active solids},
  author={Baconnier, Paul and Shohat, Dor and L{\'o}pez, C Hern{\'a}ndez and Coulais, Corentin and D{\'e}mery, Vincent and D{\"u}ring, Gustavo and Dauchot, Olivier},
  journal={Nature Physics},
  volume={18},
  number={10},
  pages={1234--1239},
  year={2022},
  publisher={Nature Publishing Group UK London}
}

@article{solon2022,
doi = {10.1088/1751-8121/ac5d82},
url = {https://dx.doi.org/10.1088/1751-8121/ac5d82},
year = {2022},
month = {apr},
publisher = {IOP Publishing},
volume = {55},
number = {18},
pages = {184002},
author = {Alexandre Solon and Jordan M Horowitz},
title = {On the Einstein relation between mobility and diffusion coefficient in an active bath},
journal = {J. Phys. A},
}

@article{maes2020,
  title = {Fluctuating Motion in an Active Environment},
  author = {Maes, Christian},
  journal = {Phys. Rev. Lett.},
  volume = {125},
  issue = {20},
  pages = {208001},
  numpages = {7},
  year = {2020},
  month = {Nov},
  publisher = {American Physical Society},
  doi = {10.1103/PhysRevLett.125.208001},
  url = {https://link.aps.org/doi/10.1103/PhysRevLett.125.208001}
}

@article{ilkerjoanny2021,
  title = {Long-time diffusion and energy transfer in polydisperse mixtures of particles with different temperatures},
  author = {Ilker, Efe and Castellana, Michele and Joanny, Jean-Fran\ifmmode \mbox{\c{c}}\else \c{c}\fi{}ois},
  journal = {Phys. Rev. Res.},
  volume = {3},
  issue = {2},
  pages = {023207},
  numpages = {7},
  year = {2021},
  month = {Jun},
  publisher = {American Physical Society},
  doi = {10.1103/PhysRevResearch.3.023207},
  url = {https://link.aps.org/doi/10.1103/PhysRevResearch.3.023207}
}

@article{grosbergjoanny2015,
  title = {Nonequilibrium statistical mechanics of mixtures of particles in contact with different thermostats},
  author = {Grosberg, A. Y. and Joanny, J.-F.},
  journal = {Phys. Rev. E},
  volume = {92},
  issue = {3},
  pages = {032118},
  numpages = {10},
  year = {2015},
  month = {Sep},
  publisher = {American Physical Society},
  doi = {10.1103/PhysRevE.92.032118},
  url = {https://link.aps.org/doi/10.1103/PhysRevE.92.032118}
}

@article{weberfrey2016,
  title = {Binary Mixtures of Particles with Different Diffusivities Demix},
  author = {Weber, Simon N. and Weber, Christoph A. and Frey, Erwin},
  journal = {Phys. Rev. Lett.},
  volume = {116},
  issue = {5},
  pages = {058301},
  numpages = {5},
  year = {2016},
  month = {Feb},
  publisher = {American Physical Society},
  doi = {10.1103/PhysRevLett.116.058301},
  url = {https://link.aps.org/doi/10.1103/PhysRevLett.116.058301}
}

@book{zwanzig2001,
  title={Nonequilibrium Statistical Mechanics},
  author={Zwanzig, R.},
  isbn={9780198032151},
  lccn={00023880},
  url={https://books.google.com/books?id=4cI5136OdoMC},
  year={2001},
  publisher={Oxford University Press}
}

@article{fkm1965,
  title={Statistical mechanics of assemblies of coupled oscillators},
  author={Ford, GW and Kac, M and Mazur, P},
  journal={Journal of Mathematical Physics},
  volume={6},
  number={4},
  pages={504--515},
  year={1965},
  publisher={AIP Publishing}
}

@article{vankampen1986,
title = {Brownian motion as a problem of eliminating fast variables},
journal = {Physica A: Statistical Mechanics and its Applications},
volume = {138},
number = {1},
pages = {231-248},
year = {1986},
issn = {0378-4371},
doi = {https://doi.org/10.1016/0378-4371(86)90183-4},
url = {https://www.sciencedirect.com/science/article/pii/0378437186901834},
author = {N.G. {Van Kampen} and I. Oppenheim}
}

@article{mori1965,
    author = {Mori, Hazime},
    title = "{Transport, Collective Motion, and Brownian Motion*)}",
    journal = {Progress of Theoretical Physics},
    volume = {33},
    number = {3},
    pages = {423-455},
    year = {1965},
    month = {03},
    issn = {0033-068X},
    doi = {10.1143/PTP.33.423},
    url = {https://doi.org/10.1143/PTP.33.423},
}

@article{feynman1963,
title = {The theory of a general quantum system interacting with a linear dissipative system},
journal = {Annals of Physics},
volume = {24},
pages = {118-173},
year = {1963},
issn = {0003-4916},
doi = {https://doi.org/10.1016/0003-4916(63)90068-X},
url = {https://www.sciencedirect.com/science/article/pii/000349166390068X},
author = {R.P Feynman and F.L Vernon},
}

@article{caldeira1983,
  title={Path integral approach to quantum Brownian motion},
  author={Caldeira, Amir O and Leggett, Anthony J},
  journal={Physica A: Statistical mechanics and its Applications},
  volume={121},
  number={3},
  pages={587--616},
  year={1983},
  publisher={Elsevier}
}

@misc{si,
      title={Statistical mechanics of a cold tracer in a hot bath}, 
      author={Amer Al-Hiyasat and Sunghan Ro and Julien Tailleur},
      year={2025},
      eprint={2503.01998},
      archivePrefix={arXiv},
      primaryClass={cond-mat.stat-mech},
      url={https://arxiv.org/abs/2503.01998}, 
}

@article{sepulveda2013collective,
  title={Collective cell motion in an epithelial sheet can be quantitatively described by a stochastic interacting particle model},
  author={Sep{\'u}lveda, N{\'e}stor and Petitjean, Laurence and Cochet, Olivier and Grasland-Mongrain, Erwan and Silberzan, Pascal and Hakim, Vincent},
  journal={PLoS computational biology},
  volume={9},
  number={3},
  pages={e1002944},
  year={2013},
  publisher={Public Library of Science San Francisco, USA}
}

@article{szamel2014self,
  title={Self-propelled particle in an external potential: Existence of an effective temperature},
  author={Szamel, Grzegorz},
  journal={Physical Review E},
  volume={90},
  number={1},
  pages={012111},
  year={2014},
  publisher={APS}
}

@article{Tailleur_2009,
doi = {10.1209/0295-5075/86/60002},
url = {https://dx.doi.org/10.1209/0295-5075/86/60002},
year = {2009},
month = {jun},
publisher = {},
volume = {86},
number = {6},
pages = {60002},
author = {J. Tailleur and M. E. Cates},
title = {Sedimentation, trapping, and rectification of dilute bacteria},
journal = {Europhysics Letters}
}

@article{sokolov2010swimming,
  title={Swimming bacteria power microscopic gears},
  author={Sokolov, Andrey and Apodaca, Mario M and Grzybowski, Bartosz A and Aranson, Igor S},
  journal={Proceedings of the National Academy of Sciences},
  volume={107},
  number={3},
  pages={969--974},
  year={2010},
  publisher={National Acad Sciences}
}

@article{di2010bacterial,
  title={Bacterial ratchet motors},
  author={Di Leonardo, Roberto and Angelani, Luca and Dell’Arciprete, Dario and Ruocco, Giancarlo and Iebba, Valerio and Schippa, Serena and Conte, Maria Pia and Mecarini, Francesco and De Angelis, Francesco and Di Fabrizio, Enzo},
  journal={Proceedings of the National Academy of Sciences},
  volume={107},
  number={21},
  pages={9541--9545},
  year={2010},
  publisher={National Acad Sciences}
}

@article{elgetiWall2013,
  title={Wall accumulation of self-propelled spheres},
  author={Elgeti, Jens and Gompper, Gerhard},
  journal={Europhysics Letters},
  volume={101},
  number={4},
  pages={48003},
  year={2013},
  publisher={IOP Publishing}
}

@article{galajda2007,
  title={A wall of funnels concentrates swimming bacteria},
  author={Galajda, Peter and Keymer, Juan and Chaikin, Paul and Austin, Robert},
  journal={Journal of bacteriology},
  volume={189},
  number={23},
  pages={8704--8707},
  year={2007},
  publisher={Am Soc Microbiol}
}

@article{Angelani_2011,
doi = {10.1209/0295-5075/96/68002},
url = {https://dx.doi.org/10.1209/0295-5075/96/68002},
year = {2011},
month = {dec},
publisher = {},
volume = {96},
number = {6},
pages = {68002},
author = {L. Angelani and A. Costanzo and R. Di Leonardo},
title = {Active ratchets},
journal = {Europhysics Letters}
}

@article{tanaka2017,
  title={Hot particles attract in a cold bath},
  author={Tanaka, Hidenori and Lee, Alpha A and Brenner, Michael P},
  journal={Physical Review Fluids},
  volume={2},
  number={4},
  pages={043103},
  year={2017},
  publisher={APS}
}

@article{caprini2020,
  title = {Spontaneous Velocity Alignment in Motility-Induced Phase Separation},
  author = {Caprini, L. and Marini Bettolo Marconi, U. and Puglisi, A.},
  journal = {Phys. Rev. Lett.},
  volume = {124},
  issue = {7},
  pages = {078001},
  numpages = {6},
  year = {2020},
  month = {Feb},
  publisher = {American Physical Society},
  doi = {10.1103/PhysRevLett.124.078001},
  url = {https://link.aps.org/doi/10.1103/PhysRevLett.124.078001}
}

@article{mipsrev2015,
  title={Motility-induced phase separation},
  author={Cates, Michael E and Tailleur, Julien},
  journal={Annu. Rev. Condens. Matter Phys.},
  volume={6},
  number={1},
  pages={219--244},
  year={2015},
  publisher={Annual Reviews}
}

@article{li2019quantifying,
  title={Quantifying dissipation using fluctuating currents},
  author={Li, Junang and Horowitz, Jordan M and Gingrich, Todd R and Fakhri, Nikta},
  journal={Nature communications},
  volume={10},
  number={1},
  pages={1666},
  year={2019},
  publisher={Nature Publishing Group UK London}
}

@article{Zia_2007,
doi = {10.1088/1742-5468/2007/07/P07012},
url = {https://dx.doi.org/10.1088/1742-5468/2007/07/P07012},
year = {2007},
month = {jul},
publisher = {},
volume = {2007},
number = {07},
pages = {P07012},
author = {R K P Zia and B Schmittmann},
title = {Probability currents as principal characteristics in the statistical mechanics of
non-equilibrium steady states},
journal = {Journal of Statistical Mechanics: Theory and Experiment},
}

@article{juelicher2007active,
  title={Active behavior of the cytoskeleton},
  author={Juelicher, Frank and Kruse, Karsten and Prost, Jacques and Joanny, J-F},
  journal={Physics reports},
  volume={449},
  number={1-3},
  pages={3--28},
  year={2007},
  publisher={Elsevier}
}

@article{sengupta2013enzyme,
  title={Enzyme molecules as nanomotors},
  author={Sengupta, Samudra and Dey, Krishna K and Muddana, Hari S and Tabouillot, Tristan and Ibele, Michael E and Butler, Peter J and Sen, Ayusman},
  journal={Journal of the American Chemical Society},
  volume={135},
  number={4},
  pages={1406--1414},
  year={2013},
  publisher={ACS Publications}
}

@article{dey2015micromotors,
  title={Micromotors powered by enzyme catalysis},
  author={Dey, Krishna K and Zhao, Xi and Tansi, Benjamin M and M{\'e}ndez-Ortiz, Wilfredo J and C{\'o}rdova-Figueroa, Ubaldo M and Golestanian, Ramin and Sen, Ayusman},
  journal={Nano letters},
  volume={15},
  number={12},
  pages={8311--8315},
  year={2015},
  publisher={ACS Publications}
}

@article{muddana2010substrate,
  title={Substrate catalysis enhances single-enzyme diffusion},
  author={Muddana, Hari S and Sengupta, Samudra and Mallouk, Thomas E and Sen, Ayusman and Butler, Peter J},
  journal={Journal of the American Chemical Society},
  volume={132},
  number={7},
  pages={2110--2111},
  year={2010},
  publisher={ACS Publications}
}

@article{baek2018generic,
  title={Generic long-range interactions between passive bodies in an active fluid},
  author={Baek, Yongjoo and Solon, Alexandre P and Xu, Xinpeng and Nikola, Nikolai and Kafri, Yariv},
  journal={Physical review letters},
  volume={120},
  number={5},
  pages={058002},
  year={2018},
  publisher={APS}
}

@article{reichhardt2017ratchet,
  title={Ratchet effects in active matter systems},
  author={Reichhardt, CJ Olson and Reichhardt, Charles},
  journal={Annual Review of Condensed Matter Physics},
  volume={8},
  number={1},
  pages={51--75},
  year={2017},
  publisher={Annual Reviews}
}

@article{julicher1997modeling,
  title={Modeling molecular motors},
  author={J{\"u}licher, Frank and Ajdari, Armand and Prost, Jacques},
  journal={Reviews of Modern Physics},
  volume={69},
  number={4},
  pages={1269},
  year={1997},
  publisher={APS}
}

@article{magnasco1993forced,
  title={Forced thermal ratchets},
  author={Magnasco, Marcelo O},
  journal={Physical Review Letters},
  volume={71},
  number={10},
  pages={1477},
  year={1993},
  publisher={APS}
}

@article{wu2000,
  title = {Particle Diffusion in a Quasi-Two-Dimensional Bacterial Bath},
  author = {Wu, Xiao-Lun and Libchaber, Albert},
  journal = {Phys. Rev. Lett.},
  volume = {84},
  issue = {13},
  pages = {3017--3020},
  numpages = {0},
  year = {2000},
  month = {Mar},
  publisher = {American Physical Society},
  doi = {10.1103/PhysRevLett.84.3017},
  url = {https://link.aps.org/doi/10.1103/PhysRevLett.84.3017}
}

@article{zhao2017enhanced,
  title={Enhanced diffusion of passive tracers in active enzyme solutions},
  author={Zhao, Xi and Dey, Krishna K and Jeganathan, Selva and Butler, Peter J and C{\'o}rdova-Figueroa, Ubaldo M and Sen, Ayusman},
  journal={Nano letters},
  volume={17},
  number={8},
  pages={4807--4812},
  year={2017},
  publisher={ACS Publications}
}

@article{ghosh2021,
   author = "Ghosh, Subhadip and Somasundar, Ambika and Sen, Ayusman",
   title = "Enzymes as Active Matter", 
   journal= "Annual Review of Condensed Matter Physics",
   year = "2021",
   volume = "12",
   number = "Volume 12, 2021",
   pages = "177-200",
   doi = "https://doi.org/10.1146/annurev-conmatphys-061020-053036",
   url = "https://www.annualreviews.org/content/journals/10.1146/annurev-conmatphys-061020-053036",
   publisher = "Annual Reviews",
   issn = "1947-5462",
   type = "Journal Article",}

@article{wilson2011differential,
  title={Differential dynamic microscopy of bacterial motility},
  author={Wilson, Laurence G and Martinez, Vincent A and Schwarz-Linek, Jana and Tailleur, J and Bryant, Gary and Pusey, PN and Poon, Wilson CK},
  journal={Physical review letters},
  volume={106},
  number={1},
  pages={018101},
  year={2011},
  publisher={APS}
}

@Article{Dunkel2010,
  author   = {Dunkel, J{\"{o}}rn and Putz, Victor B and Zaid, Irwin M and Yeomans, Julia M},
  journal  = {Soft Matter},
  title    = {{Swimmer-tracer scattering at low Reynolds number}},
  year     = {2010},
  issn     = {1744-683X},
  number   = {17},
  pages    = {4268},
  volume   = {6},
  doi      = {10.1039/c0sm00164c},
  url      = {www.rsc.org/softmatter http://xlink.rsc.org/?DOI=c0sm00164c},
}

@Article{Mino2011,
  author   = {Mi\~no, Gast\'on and Mallouk, Thomas E. and Darnige, Thierry and Hoyos, Mauricio and Dauchet, Jeremi and Dunstan, Jocelyn and Soto, Rodrigo and Wang, Yang and Rousselet, Annie and Clement, Eric},
  journal  = {Phys. Rev. Lett.},
  title    = {{Enhanced Diffusion due to Active Swimmers at a Solid Surface}},
  year     = {2011},
  issn     = {0031-9007},
  month    = {1},
  number   = {4},
  pages    = {048102},
  volume   = {106},
  arxivid  = {1012.4624},
  doi      = {10.1103/PhysRevLett.106.048102},
  url      = {https://link.aps.org/doi/10.1103/PhysRevLett.106.048102},
}

@article{Zaid2011,
author = {Zaid, Irwin M and Dunkel, J{\"{o}}rn and Yeomans, Julia M},
doi = {10.1098/rsif.2010.0545},
issn = {1742-5689},
journal = {J. R. Soc. Interface},
month = {sep},
number = {62},
pages = {1314--1331},
pmid = {21345857},
title = {{L{\'{e}}vy fluctuations and mixing in dilute suspensions of algae and bacteria}},
url = {https://royalsocietypublishing.org/doi/10.1098/rsif.2010.0545},
volume = {8},
year = {2011}
}

@Article{Foffano2012,
  author   = {Foffano, G. and Lintuvuori, J. S. and Stratford, K. and Cates, M. E. and Marenduzzo, D.},
  journal  = {Phys. Rev. Lett.},
  title    = {{Colloids in Active Fluids: Anomalous Microrheology and Negative Drag}},
  year     = {2012},
  issn     = {0031-9007},
  month    = {7},
  number   = {2},
  pages    = {028103},
  volume   = {109},
  arxivid  = {1204.1279},
  doi      = {10.1103/PhysRevLett.109.028103},
  url      = {https://link.aps.org/doi/10.1103/PhysRevLett.109.028103},
}

@Article{Mino2013,
  author    = {Mi{\~{n}}o, G. L. and Dunstan, J. and Rousselet, A. and Cl{\'{e}}ment, E. and Soto, R.},
  journal   = {J. Fluid Mech.},
  title     = {{Induced diffusion of tracers in a bacterial suspension: theory and experiments}},
  year      = {2013},
  issn      = {0022-1120},
  month     = {8},
  pages     = {423--444},
  volume    = {729},
  arxivid   = {1210.7704},
  doi       = {10.1017/jfm.2013.304},
  publisher = {Cambridge University Press},
  url       = {https://doi.org/10.1017/jfm.2013.304 https://www.cambridge.org/core/product/identifier/S0022112013003042/type/journal_article},
}

@Article{Kasyap2014,
  author  = {Kasyap, T. V. and Koch, Donald L. and Wu, Mingming},
  journal = {Phys. Fluids},
  title   = {{Hydrodynamic tracer diffusion in suspensions of swimming bacteria}},
  year    = {2014},
  issn    = {1070-6631},
  month   = {8},
  number  = {8},
  pages   = {081901},
  volume  = {26},
  doi     = {10.1063/1.4891570},
  url     = {https://doi.org/10.1063/1.4891570 http://aip.scitation.org/doi/10.1063/1.4891570},
}

@article{Morozov2014,
    title = {{Enhanced diffusion of tracer particles in dilute bacterial suspensions}},
    year = {2014},
    journal = {Soft Matter},
    author = {Morozov, Alexander and Marenduzzo, Davide},
    number = {16},
    pages = {2748},
    volume = {10},
    url = {www.rsc.org/softmatter http://xlink.rsc.org/?DOI=c3sm52201f},
    doi = {10.1039/c3sm52201f},
    issn = {1744-683X},
    pmid = {24668266}
}

@Article{Leptos2009,
  author    = {Leptos, Kyriacos C. and Guasto, Jeffrey S. and Gollub, J. P. and Pesci, Adriana I. and Goldstein, Raymond E.},
  journal   = {Phys. Rev. Lett.},
  title     = {{Dynamics of Enhanced Tracer Diffusion in Suspensions of Swimming Eukaryotic Microorganisms}},
  year      = {2009},
  issn      = {0031-9007},
  month     = {11},
  number    = {19},
  pages     = {198103},
  volume    = {103},
  doi       = {10.1103/PhysRevLett.103.198103},
  publisher = {{\`{A}}n},
  url       = {https://link.aps.org/doi/10.1103/PhysRevLett.103.198103},
}

@Article{Thiffeault2015,
  author   = {Thiffeault, Jean-Luc},
  journal  = {Phys. Rev. E},
  title    = {{Distribution of particle displacements due to swimming microorganisms}},
  year     = {2015},
  issn     = {1539-3755},
  month    = {aug},
  number   = {2},
  pages    = {023023},
  volume   = {92},
  doi      = {10.1103/PhysRevE.92.023023},
  pmid     = {26382519},
  url      = {https://link.aps.org/doi/10.1103/PhysRevE.92.023023},
}

@Article{Suma2016,
  author   = {Suma, Antonio and Cugliandolo, Leticia F. and Gonnella, Giuseppe},
  journal  = {J. Stat. Mech: Theory Exp.},
  title    = {{Tracer motion in an active dumbbell fluid}},
  year     = {2016},
  issn     = {1742-5468},
  month    = {5},
  number   = {5},
  pages    = {054029},
  volume   = {2016},
  doi      = {10.1088/1742-5468/2016/05/054029},
  url      = {https://iopscience.iop.org/article/10.1088/1742-5468/2016/05/054029},
}

@article{Kurtuldu2011,
    title = {{Enhancement of biomixing by swimming algal cells in two-dimensional films}},
    year = {2011},
    journal = {Proc. Natl. Acad. Sci. U.S.A},
    author = {Kurtuldu, Hüseyin and Guasto, Jeffrey S. and Johnson, Karl A. and Gollub, J. P.},
    number = {26},
    month = {6},
    pages = {10391--10395},
    volume = {108},
    url = {www.pnas.org/cgi/doi/10.1073/pnas.1107046108 http://www.pnas.org/cgi/doi/10.1073/pnas.1107046108},
    doi = {10.1073/pnas.1107046108},
    issn = {0027-8424},
    pmid = {21659630}
}

@article{Burkholder2017,
    title = {{Tracer diffusion in active suspensions}},
    year = {2017},
    journal = {Phys. Rev. E},
    author = {Burkholder, Eric W. and Brady, John F.},
    number = {5},
    month = {5},
    pages = {052605},
    volume = {95},
    url = {http://link.aps.org/doi/10.1103/PhysRevE.95.052605},
    doi = {10.1103/PhysRevE.95.052605},
    issn = {2470-0045},
    pmid = {28618621},
    arxivId = {1703.10554}
}

@Article{Pietzonka2018,
  author   = {Pietzonka, Patrick and Seifert, Udo},
  journal  = {J. Phys. A: Math. Theor.},
  title    = {{Entropy production of active particles and for particles in active baths}},
  year     = {2018},
  issn     = {1751-8113},
  month    = {1},
  number   = {1},
  pages    = {01LT01},
  volume   = {51},
  arxivid  = {1707.03772},
  doi      = {10.1088/1751-8121/aa91b9},
  url      = {https://doi.org/10.1088/1751-8121/aa91b9 https://iopscience.iop.org/article/10.1088/1751-8121/aa91b9},
}

@article{Kanazawa2020,
    title = {{Loopy L{\'{e}}vy flights enhance tracer diffusion in active suspensions}},
    year = {2020},
    journal = {Nature},
    author = {Kanazawa, Kiyoshi and Sano, Tomohiko G. and Cairoli, Andrea and Baule, Adrian},
    number = {7799},
    month = {3},
    pages = {364--367},
    volume = {579},
    url = {https://doi.org/10.1038/s41586-020-2086-2 http://www.nature.com/articles/s41586-020-2086-2},
    doi = {10.1038/s41586-020-2086-2},
    issn = {0028-0836},
    pmid = {32188948},
    arxivId = {1906.00608}
}

@Article{Abbaspour2021,
  author   = {Abbaspour, Leila and Klumpp, Stefan},
  journal  = {Phys. Rev. E},
  title    = {{Enhanced diffusion of a tracer particle in a lattice model of a crowded active system}},
  year     = {2021},
  issn     = {2470-0045},
  month    = {may},
  number   = {5},
  pages    = {052601},
  volume   = {103},
  doi      = {10.1103/PhysRevE.103.052601},
  pmid     = {34134202},
  url      = {https://link.aps.org/doi/10.1103/PhysRevE.103.052601},
}

@article{Reichert2021,
author = {Reichert, Julian and Granz, Leon F and Voigtmann, Thomas},
doi = {10.1140/epje/s10189-021-00039-4},
isbn = {0123456789},
issn = {1292-8941},
journal = {Eur. Phys. J. E},
month = {mar},
number = {3},
pages = {27},
title = {{Transport coefficients in dense active Brownian particle systems: mode-coupling theory and simulation results}},
url = {https://doi.org/10.1140/epje/s10189-021-00039-4 https://link.springer.com/10.1140/epje/s10189-021-00039-4},
volume = {44},
year = {2021}
}

@article{volpe2011microswimmers,
  title={Microswimmers in patterned environments},
  author={Volpe, Giovanni and Buttinoni, Ivo and Vogt, Dominik and K{\"u}mmerer, Hans-J{\"u}rgen and Bechinger, Clemens},
  journal={Soft Matter},
  volume={7},
  number={19},
  pages={8810--8815},
  year={2011},
  publisher={Royal Society of Chemistry}
}

@article{fodor2014energetics,
  title={Energetics of active fluctuations in living cells},
  author={Fodor, {\'E}tienne and Kanazawa, Kiyoshi and Hayakawa, Hisao and Visco, Paolo and Van Wijland, Fr{\'e}d{\'e}ric},
  journal={Physical Review E},
  volume={90},
  number={4},
  pages={042724},
  year={2014},
  publisher={APS}
}

@article{ben2015modeling,
  title={Modeling the dynamics of a tracer particle in an elastic active gel},
  author={Ben-Isaac, E and Fodor, {\'E} and Visco, P and Van Wijland, F and Gov, Nir S},
  journal={Physical Review E},
  volume={92},
  number={1},
  pages={012716},
  year={2015},
  publisher={APS}
}

@article{bohec2019distribution,
  title={Distribution of active forces in the cell cortex},
  author={Bohec, P and Tailleur, J and Van Wijland, F and Richert, A and Gallet, F},
  journal={Soft matter},
  volume={15},
  number={35},
  pages={6952--6966},
  year={2019},
  publisher={Royal Society of Chemistry}
}

@article{goswami2023trapped,
  title={Trapped tracer in a non-equilibrium bath: dynamics and energetics},
  author={Goswami, Koushik and Metzler, Ralf},
  journal={Soft Matter},
  volume={19},
  number={45},
  pages={8802--8819},
  year={2023},
  publisher={Royal Society of Chemistry}
}

@article{jardat2022diffusion,
  title={Diffusion of a tracer in a dense mixture of soft particles connected to different thermostats},
  author={Jardat, Marie and Dahirel, Vincent and Illien, Pierre},
  journal={Physical Review E},
  volume={106},
  number={6},
  pages={064608},
  year={2022},
  publisher={APS}
}

@article{maes1999,
  title={The fluctuation theorem as a Gibbs property},
  author={Maes, Christian},
  journal={Journal of statistical physics},
  volume={95},
  pages={367--392},
  year={1999},
  publisher={Springer}
}

@article{seifert2005,
  title={Entropy production along a stochastic trajectory and an integral fluctuation theorem},
  author={Seifert, Udo},
  journal={Physical review letters},
  volume={95},
  number={4},
  pages={040602},
  year={2005},
  publisher={APS}
}

@article{santra2023dynamical,
  title={Dynamical fluctuations of a tracer coupled to active and passive particles},
  author={Santra, Ion},
  journal={Journal of Physics: Complexity},
  volume={4},
  number={1},
  pages={015013},
  year={2023},
  publisher={IOP Publishing}
}

@inproceedings{sarkar2024harmonic,
  title={Harmonic chain driven by active Rubin bath: transport properties and steady-state correlations},
  author={Sarkar, Ritwick and Santra, Ion and Basu, Urna},
  booktitle={Proceedings A},
  volume={480},
  number={2300},
  pages={20240275},
  year={2024},
  organization={The Royal Society}
}

@article{santra2024forces,
  title={Forces from coarse-graining non-equilibrium degrees of freedom: exact results},
  author={Santra, Ion and Kr{\"u}ger, Matthias},
  journal={Journal of Statistical Mechanics: Theory and Experiment},
  volume={2024},
  number={12},
  pages={123203},
  year={2024},
  publisher={IOP Publishing}
}

@article{cui2018generalized,
  title={Generalized Langevin equation and fluctuation-dissipation theorem for particle-bath systems in external oscillating fields},
  author={Cui, Bingyu and Zaccone, Alessio},
  journal={Physical Review E},
  volume={97},
  number={6},
  pages={060102},
  year={2018},
  publisher={APS}
}

@article{pelargonio2023generalized,
  title={Generalized Langevin equation with shear flow and its fluctuation-dissipation theorems derived from a Caldeira-Leggett Hamiltonian},
  author={Pelargonio, Sara and Zaccone, Alessio},
  journal={Physical Review E},
  volume={107},
  number={6},
  pages={064102},
  year={2023},
  publisher={APS}
}

@incollection{bonetto2000fourier,
  title={Fourier's law: a challenge to theorists},
  author={Bonetto, Federico and Lebowitz, Joel L and Rey-Bellet, Luc},
  booktitle={Mathematical physics 2000},
  pages={128--150},
  year={2000},
  publisher={World Scientific}
}
\end{document}